\documentclass[twocolumn,superscriptaddress,nobalancelastpage,10pt,pra,a4paper]{revtex4}
\usepackage{amsfonts}
\usepackage{eurosym}
\usepackage{graphicx,amsmath,amssymb,amsthm}

\input{tcilatex}
\begin{document}

\title{Two--particle coined--quantum walk with long--range interaction}
\author{C\'{e}sar Alonso-Lobo, Manuel Mart\'{\i}nez-Quesada, Margarida
Hinarejos, Germ\'{a}n J. de Valc\'{a}rcel, and Eugenio Rold\'{a}n}
\affiliation{Departament d'\`{O}ptica i d'Optometria i Ci\`{e}ncies de la Visi\'{o},
Universitat de Val\`{e}ncia, Dr. Moliner 50, 46100-Burjassot}

\begin{abstract}
We present and study a two-particle quantum walk on the line in which the two
particles interact via a long-range Coulombian-like interaction. We obtain
the spectrum of the system as well as study the type of molecules that form,
attending to the bosonic or fermionic nature of the walkers. The usual loss
of distinction between attractive and repulsive forces does not entirely
apply in our model because of the long-range of the interaction.
\end{abstract}

\maketitle

\section{Introduction}

The quantum walk (QW) \cite{Feynman65} is a quantum diffusion model
re-introduced more than twenty years ago for different purposes \cite%
{Aharonov93,Meyer96,Farhi98,Childs04} and having found since then quite
diverse applications, particularly in quantum information theory \cite%
{Kendon06}. The simplest single particle QW on the line is quite well known
by today and has been the subject of recent reviews \cite{Venegas12},
including its diverse physical implementations \cite{Manouchehri14}. In this
paper we deal not with the single particle but with the two-particle QW on
the line, including a long-range interaction between the particles.

Omar et al. \cite{Omar06} were the first in considering the coined QW of two
non-interacting particles along the line, see also \cite{Sheridan06}. Their
work revealed that even in the absence of explicit interaction, the symmetry
properties of the initial state strongly affect the two-particle evolution
depending on whether the initial state is separable or entangled. In the
latter case things are different when the initial state is symmetric or
anti-symmetric under a change of indexes or, in other words, depending on
whether the walkers are of bosonic or fermionic nature if they are
indistinguishable particles; in particular, bunching (anti-bunching) is
observed for bosons (fermions). The problem put forward in \cite{Omar06} was
further studied in \cite{Stefanak06}, where the meeting problem was
addressed, and in \cite{Pathak07}, where an experimental implementation was
proposed.

The applicability of continuous two-particle QWs for solving the graph
isomorphism problem has been considered in \cite{Shiau05,Gamble10}, and
these studies have revealed that interacting QWs allow a higher
discriminating power over non-interacting QWs, a problem further studied
with coined-QWs by Berry and Wang \cite{Berry11}. In \cite{Stefanak11} the
directional correlations in two-particle coined-QWs were considered in both
the non-interacting and interacting cases, and some numerical evidence of
particle co-walking was shown in the interacting case, but it was in the
work by Albrecht et al. \cite{Ahlbrecht12} where the existence of bound
states, or molecules, in the two-particle QW with $\delta $-interaction was
demonstrated analytically, an important result that helps in understanding
the reported behaviour of these type of walks.

Interacting two-particle QWs have been studied in several contexts: the
two-particle Bose-Hubbard model \cite{Lahini12,Qin14,Beggi17,Wiater17},
which considers a nearest neighbourg interaction; in continuous-time QWs 
\cite{Peruzzo10,Poulios14,Melnikov16,Siloi16,Tang17}; and in the discrete,
coined version of the two-particle QW, both theoretically \cite%
{Stefanak06,Pathak07,Berry11,Stefanak11,Ahlbrecht12,Carson15,Wang16,Bisio18}
and experimentally \cite{Schreiber12,Sansoni12}. The evaluation of the
entanglement between the particles has received a good deal of attention. In
all cases, the considered interactions between the two particles are of
short range, nearest neighbourg in the Bose-Hubbard case and contact
interactions in the continuous- and coined-QW cases. Here we go a step
beyond by studying the coined QW of two particles that interact through a
long-range Coulombian-like interaction.

Below we demonstrate the formation of bound states by calculating the
spectrum of the system and determining their bosonic and fermionic
eigenstates. One intriguing result regarding bound states is that there
seems to be no other difference between attractive and repulsive
interactions than the sign of the quasienergy of the bound states. As
clearly stated by Albrecht et al. \cite{Ahlbrecht12} this loss of
distinction between attractive and repulsive interactions is a consequence
of the discreteness of time, which entails the loss of the distinction
between high and low energy. Indeed the formation in a periodic potential of
two body bound states with Coulomb repulsion is an obviously related
phenomenon that was predicted long ago \cite{Slater53,Hubbard63,Mahajan06}
even if observed only recently \cite{Hamo16}. In our case, however, there is
a distinction between attractive and repulsive interactions thanks to the
long-range of the interaction, and the difference between the two cases is
clearly appreciated when the two particles are far apart enough from each
other.

After this introduction the article continues with the definition of the
walk in Section II, the analysis of the spectrum in Section III, and the
analysis of the eigenstates in Sec. IV. In Sect V we discuss on the
distinction between attractive and repulsive forces, and in Section VI we
outline our main conclusions.

\section{Definition of the walk}

As in the usual two-particle QW, we consider two walkers that walk the line
by conditionally displacing to the right or left depending on their
associated qubit internal state that we denote as $\left( u_{i},d_{i}\right)$, $i=1,2$. 
A convenient four-sided coin is constructed as $\func{col}\left(
u_{1},d_{1}\right) \otimes \left( u_{2},d_{2}\right) =\func{col}\left(
u_{1}u_{2},u_{1}d_{2},d_{1}u_{2},d_{1}d_{2}\right) $, that we write as $%
\func{col}\left( r,d,u,l\right) $ by introducing the notation $r=u_{1}u_{2}$%
, $d=u_{1}d_{2}$, $u=d_{1}u_{2}$, and $l=d_{1}d_{2}$. We further consider
that the walkers interact through a Coulombian-like potential proportional
to the inverse of the walkers distance. The state of the system at
(discrete) time $t$ can be written in the form%
\begin{equation}
\left\vert \psi \right\rangle
_{t}=\sum_{c=C}\sum_{x_{1},x_{2}}C_{x_{1},t}^{x_{2}}\left\vert
x_{1},x_{2};c\right\rangle ,
\end{equation}%
where $x_{i}$ identifies the position on the line of walker $i=1,2$, $C\in
\left\{ R,D,U,L\right\} $, and $c\in \left\{ r,d,u,l\right\} $.

The state evolves as $\left\vert \psi \right\rangle _{t+1}=\hat{U}\left\vert
\psi \right\rangle _{t}$, with $\hat{U}$ a unitary that we write as 
\begin{equation}
\hat{U}=\hat{G}\hat{D}\hat{H},  \notag
\end{equation}%
being 
\begin{eqnarray}
\hat{D} &=&\sum_{x_{1},x_{2}}\hat{D}_{x_{1}x_{2}}, \\
\hat{D}_{x_{1}x_{2}} &=&\left\vert x_{1}+1,x_{2}+1;r\right\rangle
\left\langle x_{1},x_{2};r\right\vert \\
&&+\left\vert x_{1}+1,x_{2}-1;d\right\rangle \left\langle
x_{1},x_{2};d\right\vert \\
&&+\left\vert x_{1}-1,x_{2}+1;u\right\rangle \left\langle
x_{1},x_{2};u\right\vert \\
&&+\left\vert x_{1}-1,x_{2}-1;l\right\rangle \left\langle
x_{1},x_{2};l\right\vert
\end{eqnarray}%
the conditional displacement operator,%
\begin{equation}
\hat{G}=\exp \left( \frac{i\varphi }{\left\vert \hat{x}_{1}-\hat{x}%
_{2}\right\vert }\right) ,
\end{equation}%
the interaction operator, and $\hat{H}$ the coin operator. The interaction
strength is governed through the real parameter $\varphi $ in operator $\hat{%
G}$, hence $\varphi =0$ corresponds to the standard definition of a
two-particle QW along the line.

In this work we consider that the coin operator acting separately on each
qubit $\func{col}\left( u_{i},d_{i}\right) $ is the Hadamard operator 
\begin{equation}
\hat{H}_{1}=\frac{1}{\sqrt{2}}\left( 
\begin{array}{cc}
1 & 1 \\ 
1 & -1%
\end{array}%
\right) ,
\end{equation}%
with which the two--walker's coin--operator is built%
\begin{equation}
\hat{H}=\hat{H}_{1}\otimes \hat{H}_{1}=\frac{1}{2}\left( 
\begin{array}{cccc}
1 & 1 & 1 & 1 \\ 
1 & -1 & 1 & -1 \\ 
1 & 1 & -1 & -1 \\ 
1 & -1 & -1 & 1%
\end{array}%
\right) ,
\end{equation}%
which acts on vector $\func{col}\left( r,d,u,l\right) $. We note that with
this choice for the coin operator, the special case $\varphi =0$ corresponds
to the single-particle two-dimensional Hadamard walk \cite{Mackay02}.

The evolution equation can be easily put in the form of a map

\begin{subequations}
\label{mapa_xy}
\begin{eqnarray}
R_{x_{1},t+1}^{x_{2}} &=&\frac{e^{i\varphi \left\vert x_{1}-x_{2}\right\vert
^{-1}}}{2}[R_{x_{1}-1,t}^{x_{2}-1}+D_{x_{1}-1,t}^{x_{2}-1},  \notag \\
&&+U_{x_{1}-1,t}^{x_{2}-1}+L_{x_{1}-1,t}^{x_{2}-1}] \\
D_{x_{1},t+1}^{x_{2}} &=&\frac{e^{i\varphi \left\vert
x_{1}-x_{2}-2\right\vert ^{-1}}}{2}%
[R_{x_{1}-1,t}^{x_{2}+1}-D_{x_{1}-1,t}^{x_{2}+1}  \notag \\
&&+U_{x_{1}-1,t}^{x_{2}+1}-L_{x_{1}-1,t}^{x_{2}+1}], \\
U_{x_{1},t+1}^{x_{2}} &=&\frac{e^{i\varphi \left\vert
x_{1}-x_{2}+2\right\vert ^{-1}}}{2}%
[R_{x_{1}+1,t}^{x_{2}-1}+D_{x_{1}+1,t}^{x_{2}-1}  \notag \\
&&-U_{x_{1}+1,t}^{x_{2}-1}-L_{x_{1}+1,t}^{x_{2}-1}], \\
L_{x_{1},t+1}^{x_{2}} &=&\frac{e^{i\varphi \left\vert x_{1}-x_{2}\right\vert
^{-1}}}{2}[R_{x_{1}+1,t}^{x_{2}+1}-D_{x_{1}+1,t}^{x_{2}+1}  \notag \\
&&-U_{x_{1}+1,t}^{x_{2}+1}+L_{x_{1}+1,t}^{x_{2}+1}].
\end{eqnarray}%
After introducing the new indices $\rho =x_{1}-x_{2}\ $and$\ \sigma
=x_{1}+x_{2}$, the map simplifies to

\end{subequations}
\begin{subequations}
\begin{eqnarray}
R_{\rho ,t+1}^{\sigma } &=&g_{\rho }\left( R_{\rho ,t}^{\sigma -2}+D_{\rho
,t}^{\sigma -2}+U_{\rho ,t}^{\sigma -2}+L_{\rho ,t}^{\sigma -2}\right) , \\
D_{\rho ,t+1}^{\sigma } &=&g_{\rho }\left( R_{\rho -2,t}^{\sigma }-D_{\rho
-2,t}^{\sigma }+U_{\rho -2,t}^{\sigma }-L_{\rho -2,t}^{\sigma }\right) , \\
U_{\rho ,t+1}^{\sigma } &=&g_{\rho }\left( R_{\rho +2,t}^{\sigma }+D_{\rho
+2,t}^{\sigma }-U_{\rho +2,t}^{\sigma }-L_{\rho +2,t}^{\sigma }\right) , \\
L_{\rho ,t+1}^{\sigma } &=&g_{\rho }\left( R_{\rho ,t}^{\sigma +2}-D_{\rho
,t}^{\sigma +2}-U_{\rho ,t}^{\sigma +2}+L_{\rho ,t}^{\sigma +2}\right) .
\end{eqnarray}%
with 
\end{subequations}
\begin{equation}
g_{\rho }=\frac{1}{2}\exp \left( \frac{i\varphi }{\left\vert \rho
\right\vert }\right) ,  \label{gr}
\end{equation}%
the interaction coupling between the two particles.

The probability of finding the walkers at positions $\left( \sigma ,\rho
\right) $ at time $t$ is given by $P_{\rho ,t}^{\sigma }=\sum_{C}\left\vert
C_{\rho ,t}^{\sigma }\right\vert ^{2}$, with $C\in \left\{ R,D,U,L\right\} $%
. Below we show results for $P_{x_{1},t}^{x_{2}}$, and also for the marginal
probabilities $P_{\rho ,t}=\sum_{\sigma }P_{\rho ,t}^{\sigma }$ and $%
P_{\sigma ,t}=\sum_{\rho }P_{\rho ,t}^{\sigma }$.

\section{Spectrum}

Notice first that the interaction coupling just depends on the modulus of
the relative coordinate, $\left\vert \rho \right\vert $; hence one can look
for solutions of the form%
\begin{equation}
C_{\rho ,t}^{\sigma }=e^{i\left( \omega t-k\sigma \right) }C_{\rho },
\end{equation}%
with $C\in \left\{ R,D,U,L\right\} $, which are plane waves propagating
along coordinate $\sigma =x_{1}+x_{2}$ with pseudo-energy $\omega \in \left[
-\pi ,+\pi \right[ $ and pseudo-momentum $k\in \left[ -\pi /2,+\pi /2\right[ 
$. After substitution, one easily gets

\begin{subequations}
\label{mapa_int}
\begin{eqnarray}
e^{i\omega }R_{\rho } &=&g_{\rho }e^{i2k}\left( R_{\rho }+D_{\rho }+U_{\rho
}+L_{\rho }\right) , \\
e^{i\omega }D_{\rho } &=&g_{\rho }\left( R_{\rho -2}-D_{\rho -2}+U_{\rho
-2}-L_{\rho -2}\right) , \\
e^{i\omega }U_{\rho } &=&g_{\rho }\left( R_{\rho +2}+D_{\rho +2}-U_{\rho
+2}-L_{\rho +2}\right) , \\
e^{i\omega }L_{\rho } &=&g_{\rho }e^{-i2k}\left( R_{\rho }-D_{\rho }-U_{\rho
}+L_{\rho }\right) ,
\end{eqnarray}%
which can be numerically diagonalized, so that the pseudo-energy spectrum
and bound states can be obtained.

The map above is invariant under the swaping $\left\{ k,R,U\right\}
\leftrightarrow \left\{ -k,L,-D\right\} $. Moreover, the change $\varphi
\rightarrow -\varphi $ provides the map corresponding to plane-waves $%
e^{-i\left( \omega t-k\sigma \right) }$, so that passing from an attractive
to a repelling potential consists in changing $\left( \omega ,k\right) $ by $%
\left( -\omega ,-k\right) $. Hence, when molecules are formed they will form
irrespective of the interaction sign, with the only difference that
quasi-energy signs are reversed. This symmetry makes unnecessary any further
discussion about the influence of the interaction potential sign in the
spectrum, so that we take it to be positive in what follows.

A second consequence from the map form comes from the fact that first
neighbourg sites are not coupled in coordinate $\rho $ (notice that the
connected sites are $\rho \leftrightarrow \rho \pm 2$), which makes the
problem is separable into two, for $\rho $ even and odd, depending on the
initial condition, i.e., for a localized initial state in which the two
particles start at the same (adjacent) positions, $\rho $ will always be
even (odd). A major difference between the two cases is that for odd $\rho $
we do not need to give a value to $g_{0}$, while for even $\rho $ one must
fix the self-energy as $g_{0}$ is not well defined, see (\ref{gr}).
\begin{figure}
[ptb]
\includegraphics[scale=0.43]{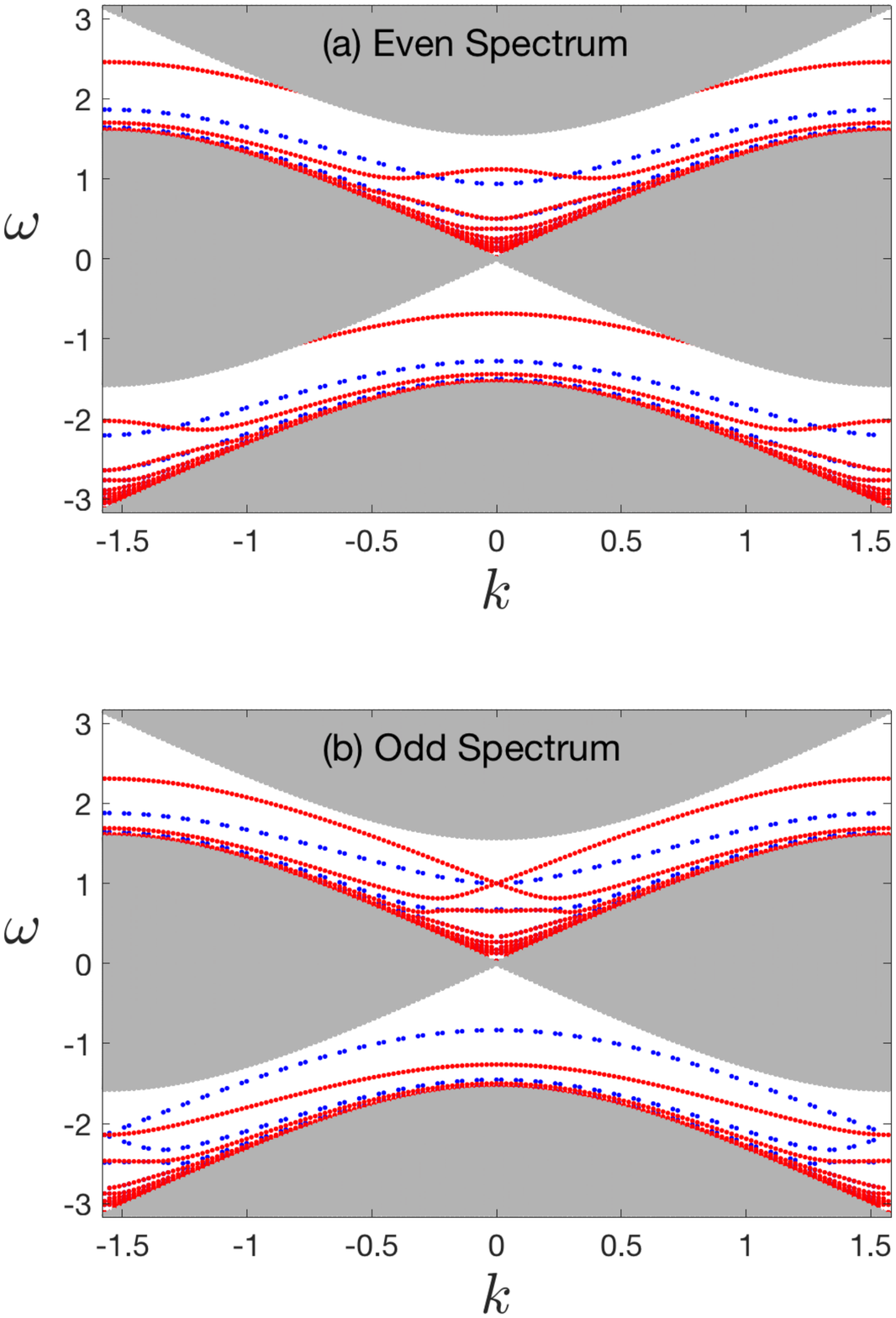} \caption{Spectrum of the two-particle walk for $%
\protect\varphi =1$ and a circle with periodic boundary conditions and
lengths $l_{c}=190$ and $l_{c}=191$ for the even and odd spectra,
respectively. In the latter, $\protect\varphi _{0}=2\protect\pi $. The
grey-shadowed area corresponds to the continuous part of the spectrum, the
blue-dashed (red-continuous) lines correspond to fermions (bosons).}
\end{figure}

In Fig. 1 we show the numerically obtained spectra for $\varphi =1$ in both
the even and odd cases, Figs. 1(a) and 1(b) respectively. In solving (\ref%
{mapa_int}), we have imposed periodic boundary conditions on a line of
length $l_{c}=190$ for the even case and $l_{c}=191$ for the odd case, the
lengths having been chosen large in order to capture the behaviour in the
continuum and also in order to capture not only small size bound states but
also large ones. For the even case we have taken $g_{0}=e^{i\varphi _{0}}/2$
with $\varphi _{0}=\pi /2$. The figures reveal the existence of both a
continuous spectrum (represented as a grey shadowed area) and a discrete
spectrum [the blue-dashed (red-continuous) lines corresponding to fermions
(bosons), which we study in the following section]. The continuous part
corresponds to plane waves along both coordinates $\rho $ and $\sigma $, and
its analytical expression is easy to derive in the limit $l_{c}\rightarrow
\infty $ (see \cite{Ahlbrecht12} where it is derived and represented).
\begin{figure}
[ptb]
\includegraphics[scale=0.43]{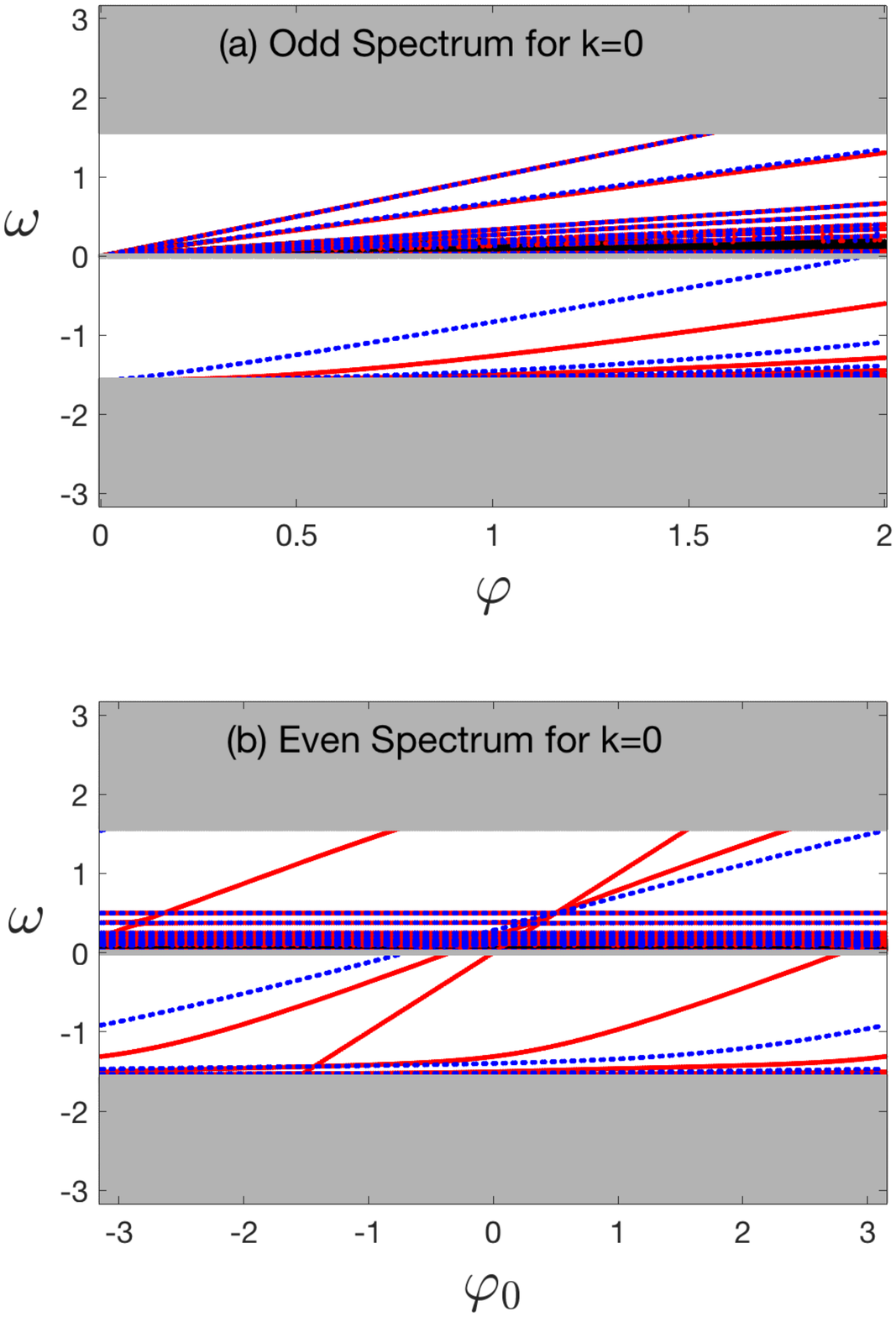} \caption{Spectra of the two-particle walk for $%
\protect\varphi =1$ and $k=0$. In (a) the dependence with $\protect\varphi $
is shown for the odd case, and in (b) the dependence with $\protect\varphi %
_{0}$ for $\protect\varphi =1$ in the even case is shown. The color code and
the rest of details as in Fig. 1.}
\end{figure}

In Figs. 2 the influence of $\varphi $ for fixed $k=0$ in the odd case, Fig.
2(a), as well as the influence in the even case of $\varphi _{0}$ for fixed $%
\varphi =1$ and $k=0$, Fig. 2(b), are shown. The increase of $\varphi $
rapidly increases the complexity of the discrete spectrum richness. As for
the role of $\varphi _{0}$, it consists in shifting the energy of a part of
the eigenstates, not of all of them; we can conclude that the states whose
energy is unaffected by $\varphi _{0}$ are those that have null (or nearly
null) projection on position $\rho =0$, while it strongly affects states
with large projection on $\rho =0$. Another important feature that Fig. 2
reveals is that the bound states energy moves into de continuum as $\varphi $
or $\varphi _{0}$ are changed, which means that in the continuum part of the
spectrum there are not only plane waves but also bound states.

\section{Bound states}

The numerical diagonalization of map (\ref{mapa_int}) provides the complete
set of eigenstates. It must be taken into account that, for
indistinguishable particles, the wavefunction must verify to be either
symmetric or antisymmetric under a change of the particle index, which
correspond to bosons and fermions, respectively. In our case, given the
definitions $r=u_{1}u_{2}$, $d=u_{1}d_{2}$, $u=d_{1}u_{2}$, $l=d_{1}d_{2}$,
the change $1\longleftrightarrow 2$ implies the changes $\left( R_{\rho
},D_{\rho },U_{\rho },L_{\rho }\right) \rightarrow \pm \left( R_{-\rho
},U_{-\rho },D_{-\rho },L_{-\rho }\right) $ for bosons ($+$) and fermions ($%
- $).

There is a class of bound states that can be obtained analytically in a very
simple manner. These bound sates are those in which the two particles remain
at a fixed distance $\rho _{0}$ with a null probability of being at any
other distance $\rho \neq \rho _{0}$. Consider first the odd case; take in (%
\ref{mapa_int}) all the amplitudes null but those for $\rho =\pm 1$, for $%
k=0 $, and the result is $\omega =\varphi $ and 
\end{subequations}
\begin{eqnarray}
U_{+1} &=&D_{-1}=0,\ \ U_{-1}=D_{+1},  \notag \\
L_{\pm 1} &=&R_{\pm 1}-D_{+1},
\end{eqnarray}%
plus the normalization condition. As there are three degrees of freedom to
fix the state, the resulting state is triply degenerated. Three possible
states are those for which (i) $L_{+1}=R_{+1}$, $L_{-1}=R_{-1}=0$, and $%
U_{-1}=D_{+1}=0$ (the molecule only occupies $\rho =+1$), (ii) the same but
with $L_{-1}=R_{-1}$ and $L_{+1}=R_{+1}=0$ (the molecule only occupies $\rho
=-1$), and (iii) $U_{-1}=D_{+1}\neq 0$ (the molecule occupies both positions 
$\rho =\pm 1$). Neither of these bound states cannot be classified as boson
nor fermion, but from (i) and (ii)\ above (i.e. $L_{+1}=R_{+1}$ and $%
L_{-1}=R_{-1}$) one can construct a boson or fermion by choosing $%
L_{+1}=L_{-1}$ or $L_{+1}=-L_{-1}$, respectively; and state (iii)\ can be
chosen to be a boson by taking $L_{+1}=L_{-1}$. Indeed, as $k$ becomes non
null, the degeneracy breaks with two of the states corresponding to fermions
and one to a boson, Fig. 1(b).

By assuming that all the amplitudes are null but those for $\rho =\rho _{0}$
with $\left\vert \rho _{0}\right\vert \geqslant 3$, for $k=0$ one obtains
that the pseudo-energy goes like $\omega =\varphi /\left\vert \rho
_{0}\right\vert $ and the state amplitudes as%
\begin{eqnarray}
U_{\rho _{0}} &=&D_{\rho _{0}}=0,  \notag \\
L_{\rho _{0}} &=&R_{\rho _{0}},
\end{eqnarray}%
this state being degenerate because $\rho _{0}$ can be either positive or
negative. Again, by combining this state with the equivalent one existing at 
$-\rho _{0}$, a boson and a fermion are constructed. All these bound states
are easily identified in the spectrum of Fig. 2(a) because the pseudo-energy
goes like $\omega =\varphi /\left\vert \rho _{0}\right\vert $.

In the even case, a similar reasoning permits to find several states easily,
the simplest being the state of energy $\omega =\varphi _{0}$ when $\rho =0$
and states of energy $\omega =\varphi /\rho _{0}$ when $\left\vert \rho
_{0}\right\vert \geqslant 2$. Notice that all the bound states we have
commented, both in the odd and even cases, are degenerated at $\omega =0$
when $\varphi =0$, see Fig. 2(a), and it is the interaction that breaks this
degeneracy. Of course these are not the only bound states, only the simpler.
%
\begin{figure}
[ptb]
\includegraphics[scale=0.47]{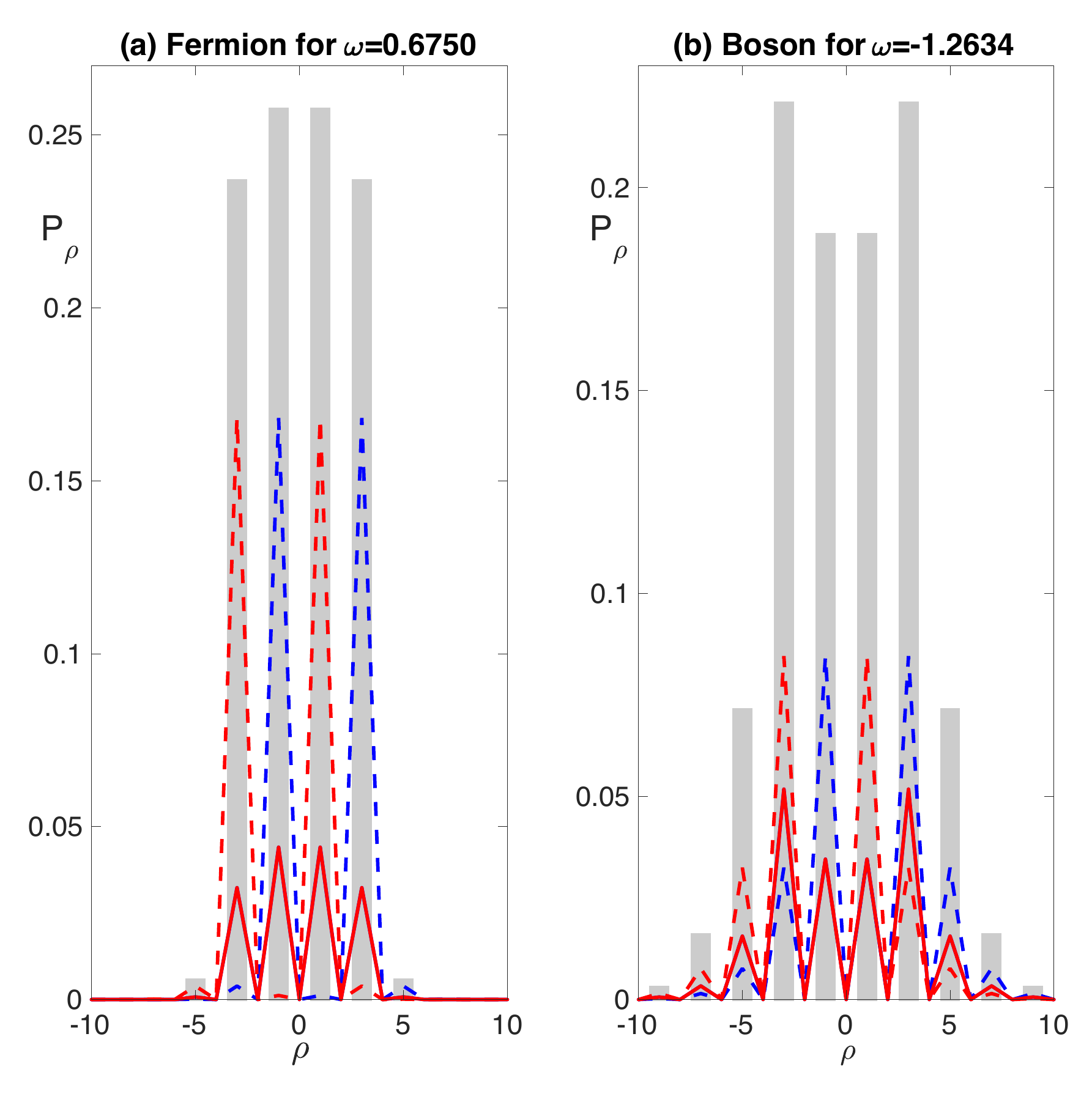} \caption{Bound states in the odd case
for $\protect\varphi =1$, $k=0$, and $\protect\omega =0.6750$ (left) and $%
\protect\omega =-1.2634$ (right), which are a fermion and a boson,
respectively. The grey bars mark $P_{\protect\rho }$, the total probability
as a function of the relative coordinate $\protect\rho $. The dashed-blue
and dashed-red lines corresponds to $\left\vert R\right\vert ^{2}$ and $%
\left\vert L\right\vert ^{2}$, respectively, while the full-red line
corresponds to $\left\vert D\right\vert ^{2}=\left\vert U\right\vert ^{2}$.}
\end{figure}
Apart from the simple bound states just described, the spectrum reveals the
existence of other more complex states extending over several positions
along $\rho $, an example of which is given in Fig. 3 where the bound state
detailed probability distribution is shown as a function of $\rho $, for
both a boson and a fermion, see the figure caption for details. In order to
give a feeling of the different bound states, we show in Fig. 4 the total
probability distribution as a function of $\rho $ for $\varphi =1$ and $k=0$%
. The plot shows that most molecules can be calculated by diagonalizing the
map on a line with a small total length $l_{c}$, so that our choice $%
l_{c}=190,191$ is more than correct.
\begin{figure}
[ptb]
\includegraphics[scale=0.22]{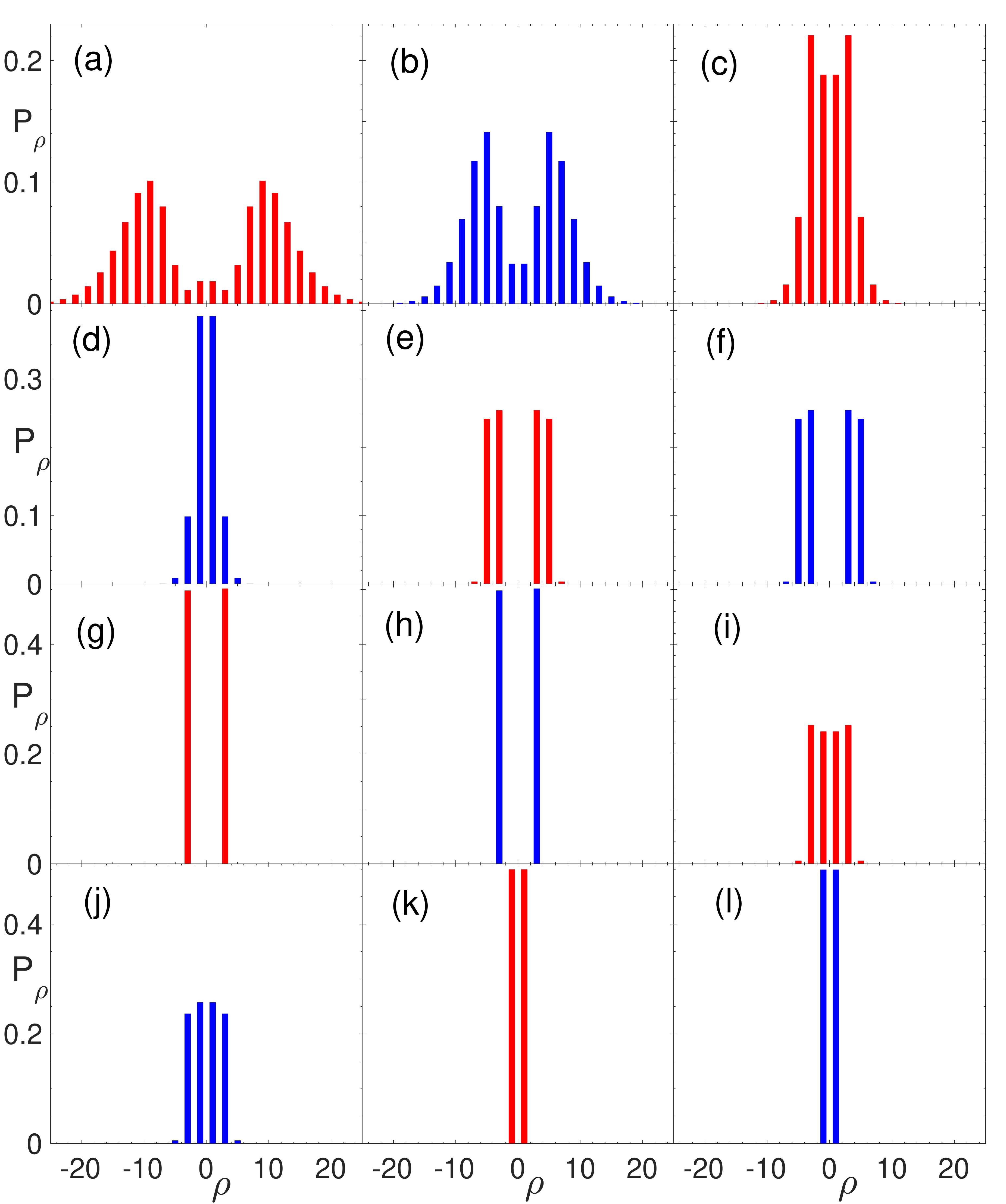} \caption{Total probability as a function of $\protect\rho $ for different bound
states corresponding to the even case with $\protect\varphi =1$ and $k=0$.
The red (blue) graphs correspond to ferminons (bosons), and the states are
represented for increasing energies, the frequencies of the states being:
(a) -1.499, (b) -1.453, (c) -1.2631, (d) -0.8328, (e) 0.2677, (f) 0.2678,
(g) 0.333 (double), (h) 0.333 (double), (i) 0.6546, (j) 0.6750, (k) 1
(triple), (l) 1 (triple). Bound states (c) and (j) were shown in full detail
in Fig. 3 above. Notice the different size of the vertical axes in the
different rows.}
\end{figure}
Ahlbrecht et al. \cite{Ahlbrecht12} were able to demonstrate that the bound
states follow a joint QW on the line. Here we shall not attempt any
analytical approach to that but just show numerically that this is the case
also in our problem. In order to project into a particular bound state we
choose as an initial condition that the two walkers lie in molecules (k) and
(l) in Fig. 5, one is a boson and the other is a fermion; further, this
initial condition occupies odd positions from $-13$ to $+13$ along the $%
\sigma $ coordinate (14 occupied sites in total). We choose such extended
initial states in order to project most of the probability distribution onto
the selected bound states. When the QW is run we clearly see that most of
the probability lies on the bound state (see the insets showing that the
probability mostly occupies $\rho $ positions corresponding to the bound
state, extending very little along this coordinate), and that these bound
states behave as expected from the spectrum of Fig. 1(b): notice that at
around $k=0$ and $\omega =1$ to the fermion it corresponds a parabolic
dispersion relation (that increases the probability distribution width with
time, see Fig. 5(b)), while to the boson there correspond two dispersion
relations that give a mean velocity to the distribution, see Fig. 5(a),
where two wavepackets are seen to travel in opposite directions, each
governed by a dispersion relation with a different sign of the velocity.
This example shows how to every molecule there corresponds different
propagation properties.

\begin{figure}
[ptb]
\includegraphics[scale=0.55]{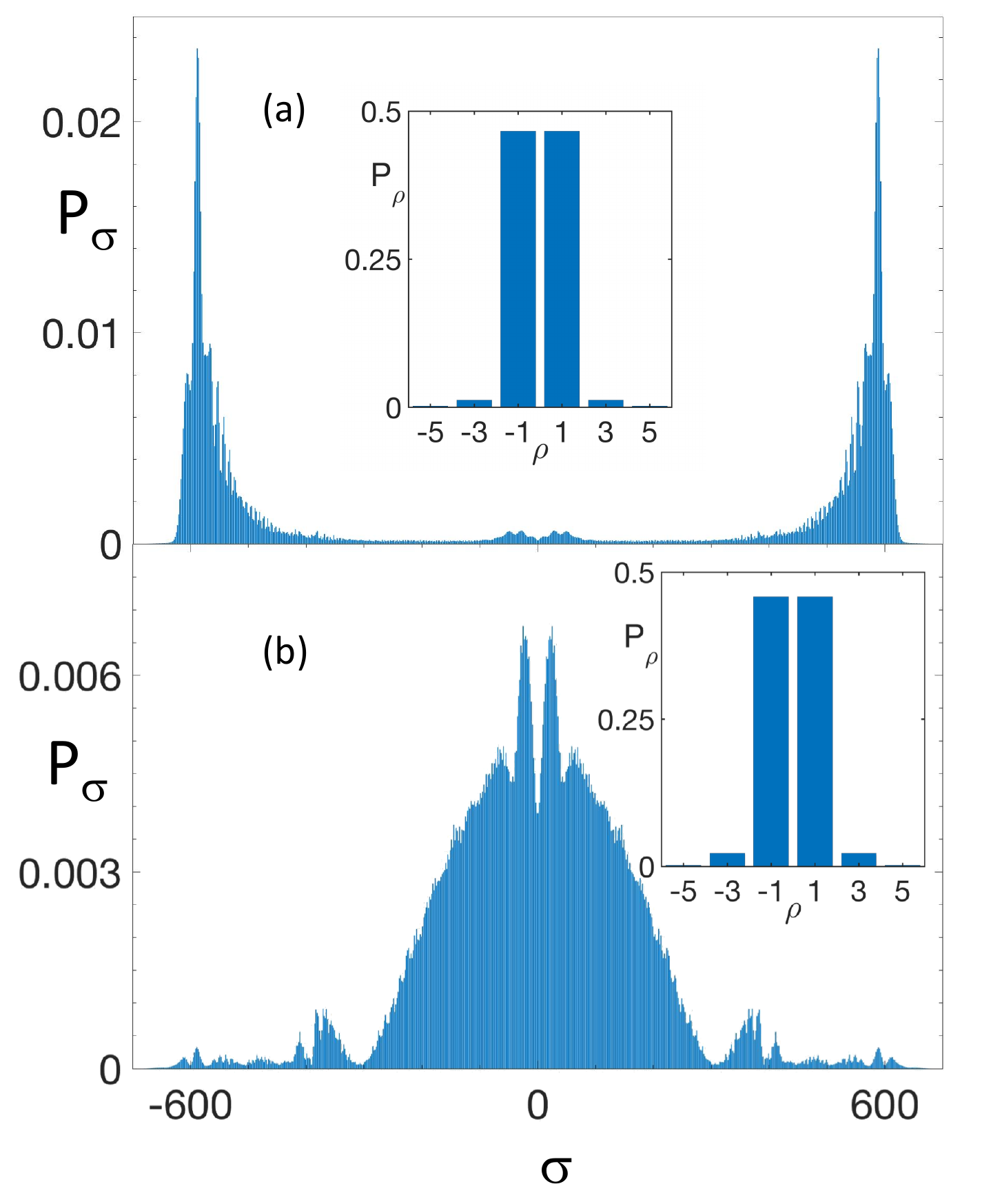} \caption{Marginal probability distribution $P_{\protect\sigma }$ versus coordinate $%
\protect\sigma $ for $\protect\varphi =1$ at $t=500$ corresponding to the
odd case; the insets show the marginal probability $P_{\protect\rho }$. The
initial coin state is $\left( 0,0,1/\protect\sqrt{2},1/\protect\sqrt{2}%
\right) $, i.e., $\left( 0,1\right) $ for the first coin and $\left( 1/%
\protect\sqrt{2},1/\protect\sqrt{2}\right) $ for the second; as for the
initial condition, it has been chosen to occupy odd positions from $-13$ to $%
+13$ along the $\protect\sigma $ coordinate (14 ocupied sites in total) with
uniform amplitude. The bound states corresponding to (a)\ are bosons, while
in (b) the bound state is a fermion. }
\end{figure}

\section{Attractive vs repulsive interactions}

One interesting question is whether there is any clear difference between
attractive and repulsive interactions, and certainly there should be if one
expects the long-wavelength approximation of the Coulombian QW to capture
some of the physics of Coulombian interactions in the continuum. We have
seen that as far as the spectrum is involved, the only difference between
the two cases consists in changing the pseudo-energy sign. However, by
taking a initial condition that does not project on any of the bound-states,
which is accomplished by taking a large initial distance between the
walkers, one expects to see the difference between attraction and repulsion
because of the long range of the interaction. And this is exactly what
numerics show: a transient attraction/repulsion between the two walkers that
lasts till the diffusion leads to the spatial overlap of the probability
distribution of the two walkers, from this time on the subsequent evolution
of the probability distribution being no more governed by the Coulombian
attraction/repulsion but by the projection onto the different bound states.

In Figs.6 and 7 we show, for attractive and repulsive interactions, the
temporal evolution of the mean distance between the particles $\left\langle
\rho \right\rangle $ minus the initial distance $\rho _{0}$. They correspond
to two distinguishable walkers whose initial positions are centered at $%
x_{1,2}\left( t=0\right) =\pm 300$, hence $\rho _{0}=600$, the walkers not
being sharply localized but extending along the line with a Gaussian
distribution of moderate width $\Delta x=10$. The insets in the figures show
top views of the probability distributions $P_{x_{1},t}^{x_{2}}$ on the
plane $\left( x_{1},x_{2}\right) $ at three selected instants (bright
colours correspond to large probability values).

\begin{figure}
[ptb]
\includegraphics[scale=0.27]{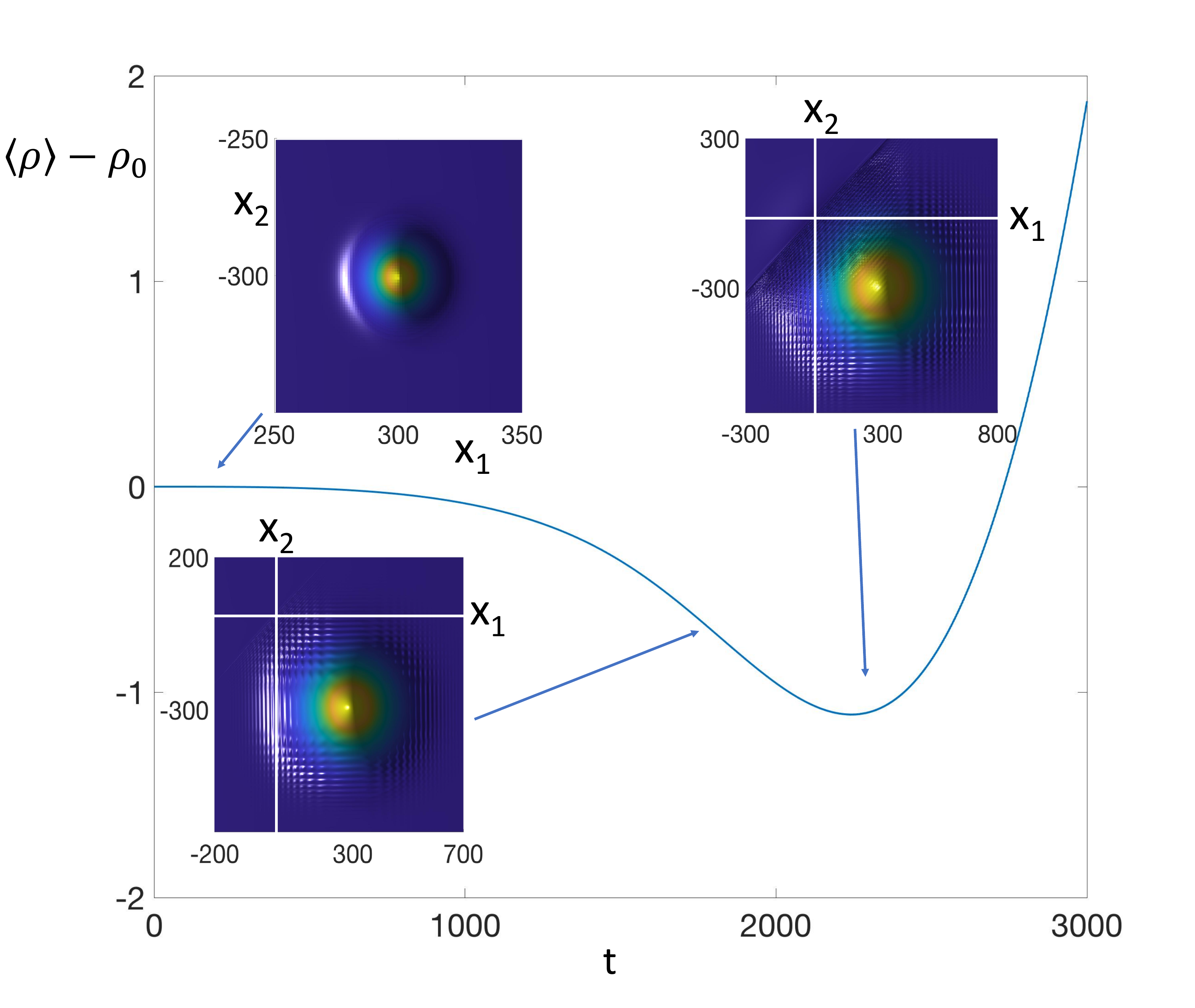} \caption{Mean value of the distance between the particles $%
\left\langle \protect\rho \right\rangle $, referred to the initial distance $%
\protect\rho _{0}=600$, for two particles initally located at $x_{1,2}\left(
t=0\right) =\pm 300$ both having an initial Gaussian distribution of width $%
\Delta x=10$ (\textrm{FWHM}$=20\ln 2$) and $k_{x}=k_{y}=\protect\pi /4$. The
initial coin state is $\left( 1,i,-i,1\right) /2$, i.e., $\left( 1,-i\right)
/\protect\sqrt{2}$ for the first coin and $\left( 1,i\right) /\protect\sqrt{2%
}$ for the second. The insets show a top-view of the probability
distribution $P_{x_{1},t}^{x_{2}}$ on the plane $\left( x_{1},x_{2}\right) $
(clear colors correspond to large probability). Notice that the two
distributions do not project on each other initially (in the left inset the
axes are far apart), the start projecting in the middle inset (the total
distribution approaches the origin), and begin to project strongly on the
right inset. We took $\protect\varphi =\protect\pi $. }
\end{figure}
If there were no interaction, diffusion would not make the two particles
distribution overlap each other before some time because they are far apart
initially, and during these initial stages attraction/repulsion can be
captured, as Figs. 6 and 7 show. When diffusion finally makes that the
probability distributions mix [this occurs when the total probability
distribution reaches the diagonal on the plane $\left( x_{1},x_{2}\right) $,
see the insets in Figs. 6 and 7], the subsequent dynamics is governed by
that of the bound and unbound states onto which the probability distribution
projects, which strongly depends on the initial condition. In order to make
these calculations we first looked at the dispersion relation of the
non-interacting walk \cite{Ahlbrecht12}, and concluded that by taking $%
k_{x}=k_{y}=\pi /2$ the evolution for $\varphi =0$ is that corresponding to
the usual dispersion of a Gaussian wave-packet \cite{deValcarcel10} (the
wavepacket width increases slowly with time without net motion), which is
the ideal situation to isolate the effect of the interaction between
particles. But this is not essential, and if one takes different initial
values for $k_{x}$ and $k_{y}$, one obtains similar results for the
evolution of the distance between the centroids of the probability
distributions, even if it may split into several wavepakets. Moreover, the
above qualitative conclusion applies equally well when the initial condition
is point like, the major difference being that the duration of the
Coulombian transient is shorter in this case because of the faster spread of
the probability distribution for localized initial states.

\begin{figure}
[ptb]
\includegraphics[scale=0.27]{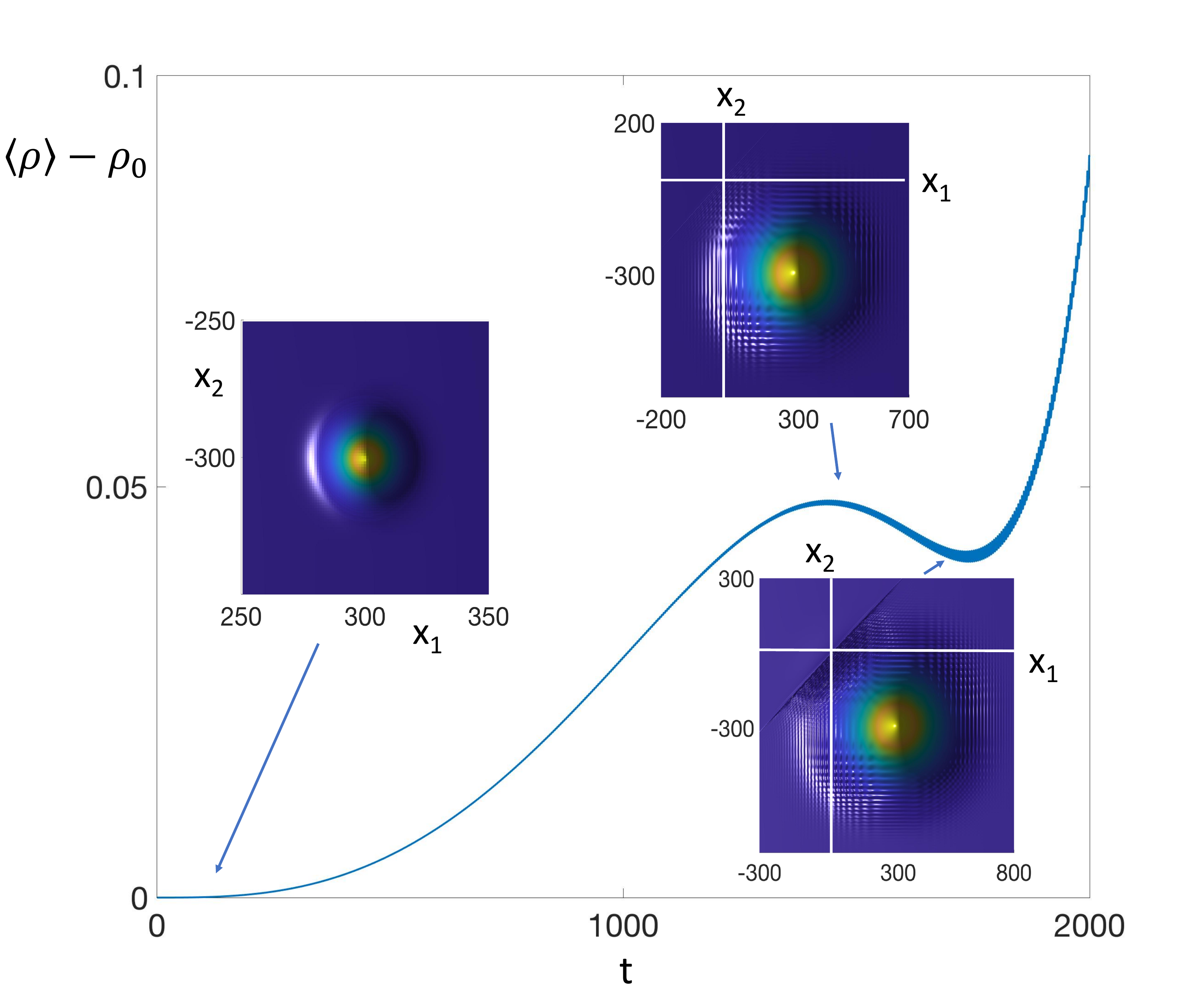} \caption{As in Fig. 6 but for $\protect\varphi =-\protect%
\pi $.}
\end{figure}
When indistinguishable particles are considered, only the repulsion can be
studied as the two particles must have the same charge, and we have tested
that the same qualitative conclusions hold.

\section{Conclusions}

In this article we have introduced the QW of two particles that interact via
a long-range Coulombian-like interaction. The novelty with respect to
previous work on two-particle QWs lies on the long-range of the interaction
as contact interactions are the only ones studied up to know in coined QWs,
moreover, in the wider context of continuos-time QWs and Bose-Hubbard
two-particle models next-neighbourg or contact interactions are the only
ones considered. We have been able to obtain the spectrum of plane waves as
well as to derive the bound states, both bosons and fermions, of which we
have given some details. We have also shown, numerically, that the bound
states follow a joint QW that is governed by the dispersion relation
corresponding to those bound states. We have finally discussed on the
differences between attractive and repulsive interactions, that better
manifest by studying the transient dynamics when the particles lie initially
apart enough.

We acknowledge financial support from the Spanish Government and the
European Union FEDER through project FIS2014-60715-P.

\end{document}